\documentclass[aps,prc,preprint,superscriptaddress]{revtex4-1}

\usepackage{graphicx}
\usepackage{amsmath,bm,ascmac,amssymb,calc}
\usepackage{CJK}

\begin{document}
\begin{CJK*}{UTF8}{}
\CJKfamily{min}


\title{Significance of tensor force in pseudo-spin symmetry}


\author{H. Nakada}
\email[E-mail:\,\,]{nakada@faculty.chiba-u.jp}
\affiliation{Department of Physics, Graduate School of Science,
  Chiba University,\\
  Yayoi-cho 1-33, Inage, Chiba 263-8522, Japan}
\affiliation{Research Center for Nuclear Physics, Osaka University,\\
  Mihogaoka 10-1, Ibaraki, Osaka 567-0047, Japan}

\author{T. Inakura}
\email[E-mail:\,\,]{inakura@gmail.com}
\affiliation{Office of Institutional Research and Decision Support,
  Tokyo Institute of Technology,\\
  Meguro, Tokyo 152-8550, Japan}
\affiliation{Laboratory for Zero-Carbon Energy, Institute of Innovative Research,
  Tokyo Institute of Technology,\\
  Meguro, Tokyo 152-8550, Japan}


\date{\today}

\begin{abstract}
  We point out that the pseudo-spin symmetry (PSS) of nuclei
  significantly depends on the proton ($Z$) and neutron numbers ($N$),
  sometimes giving rise to the characteristic structures.
  By using the non-relativistic spherical Hartree-Fock calculation
  with a realistic tensor force,
  we show that the tensor force may be deeply relevant
  to the $Z$- and $N$-dependence of the PSS.
  While the PSS has often been discussed
  in the context of the relativistic symmetry,
  the tensor-force effects on the PSS
  sometimes look analogous to the $Z$-and $N$-dependence of the PSS
  in the relativistic mean-field (RMF) calculations
  without explicit tensor force.
  The observed variation of the $p0d_{3/2}$-$p1s_{1/2}$ levels
  from $^{40}$Ca to $^{34}$Si
  is consistent with the tensor-force-driven $Z$-dependence of the PSS,
  but not necessarily with the RMF result.
  Even though it is too early to be conclusive,
  this result elucidates the significance of the tensor force
  when discussing the PSS.
\end{abstract}


\maketitle
\end{CJK*}



\textit{Introduction.}\quad

Various symmetries have been found and discussed in nuclear structure physics.
The pseudo-spin symmetry (PSS)~\cite{ref:AHS69,ref:HA69} is among them,
being crucial for some characteristic structures as discussed below.
The so-called pseudo-LS coupling scheme and the pseudo-SU(3) model
were also proposed on top of the PSS~\cite{ref:AHS69}.
In short, the PSS is the near degeneracy between $(n,\ell,j=\ell+1/2)$
and $(n'=n-1,\ell'=\ell+2,j'=\ell'-1/2)$ single-particle (s.p.) orbitals
found in a certain region of the nuclear chart,
where $(n,\ell,j)$ stands for the radial, orbital, and summed angular-momentum
quantum numbers.
For those pairs,
the pseudo-radial and orbital quantum numbers are assigned
as $\tilde{n}=n-1=n'$ and $\tilde{\ell}=\ell+1=\ell'-1$.
It has been argued since the late 1990s
that the PSS is a relativistic symmetry~\cite{ref:Gin97,ref:Men98,ref:LMZ15},
as $\tilde{\ell}$ is the real orbital angular momentum
of the lower component of the Dirac spinor,
rather than a hypothetical quantum number.

The tensor force has been pointed out to give rise to
proton- ($Z$) and neutron-number ($N$) dependence
of the shell structure~\cite{ref:Vtn,ref:Les07,ref:SC14,ref:NS14},
which is sometimes called ``shell evolution''.
This $Z$- and $N$-dependence of the shell structure
should be relevant to the PSS.
Whereas tensor-force effects on the PSS have been argued~\cite{ref:WYLG13},
sufficient attention has not been paid to the $Z$- and $N$-dependence.
In this paper,
we discuss how the tensor force affects the PSS,
with particular interest in the variation
due to the occupation of specific orbits.

We apply the spherical Hartree-Fock (HF) calculations,
and compare the s.p. energy spacings between the PSS partners
among effective interactions (or energy-density functionals)
with and without the tensor force.
Because the tensor force is active in $jj$-closed configurations
while inactive under the spin saturation~\cite{ref:SNM16},
the variation of the spacings between a $\ell s$-closed nucleus
and a $jj$-closed one provides a good indication of the tensor-force effects.
In investigating the tensor-force effects,
the M3Y-type semi-realistic interaction~\cite{ref:Nak03}
has a desirable property.
It contains a realistic tensor force,
which has been derived from a bare nucleonic interaction
through the $G$-matrix~\cite{ref:M3Y-P} without further adjustment,
and at the same time the strength of the tensor force
is well examined in the shell structure~\cite{ref:NSM13}.
We calculate s.p. energies with the M3Y-P6 interaction~\cite{
  ref:Nak13,ref:Nak20},
which provides the magic numbers compatible with most experimental data
up to unstable nuclei~\cite{ref:NS14}.
They are compared with those of the Gogny-D1S interaction~\cite{ref:D1S},
which is a representative of the phenomenological interactions
that have been used for a variety of nuclear structure studies
with no explicit tensor force.
To clarify the tensor-force effects,
we also present the s.p. energies obtained with M3Y-P6
but after subtracting the contribution of the tensor force.
The s.p. levels in the relativistic mean-field (RMF) approach
with the DD-ME2 parameter-set~\cite{ref:DD-ME2} are shown as well,
for which the DIRHB code~\cite{ref:DIRHB} has been employed.
Note that, under the time-reversal symmetry,
the tensor force has contributions only from the exchange terms~\cite{
  ref:SNM16,ref:VB72,ref:WZLL18},
which are not included in the RMF calculations.
Experimental information on the s.p. levels
can be obtained from the low-lying levels adjacent to the doubly-closed nuclei.
Despite the fragmentation of the s.p. strengths,
the energy difference between the lowest-lying states
is often an acceptable measure,
particularly within a single nuclide.
We shall compare the calculated levels with the available experimental results.

\textit{$Z$- and $N$-dependence of pseudo-spin symmetry.}\quad

While the PSS plays important roles in nuclear structure,
we here point out that the PSS has $Z$- and $N$-dependence,
well maintained in certain regions but almost lost in others.
In discussing the origin of the PSS,
the $Z$- and $N$-dependence should not be discarded.

At the mean-field level,
the tensor force has the following effects~\cite{ref:Vtn,ref:Les07,ref:SC14,
  ref:NS14,ref:SNM16,ref:Nak20}.
It predominantly acts between a proton and a neutron.
At the $\ell s$ closure, the tensor-force effects become negligibly small
owing to the spin saturation.
When a proton (neutron) orbit with $j=\ell+1/2$ is occupied,
the tensor force pushes up neutron (proton) $j'=\ell'+1/2$ orbits
and pushes down neutron (proton) $j'=\ell'-1/2$ orbits.
Thus, the tensor force yields significant $Z$- and $N$-dependence
of the s.p. level spacings
through the occupation of specific orbitals.

\begin{figure}
  \includegraphics[scale=1.0]{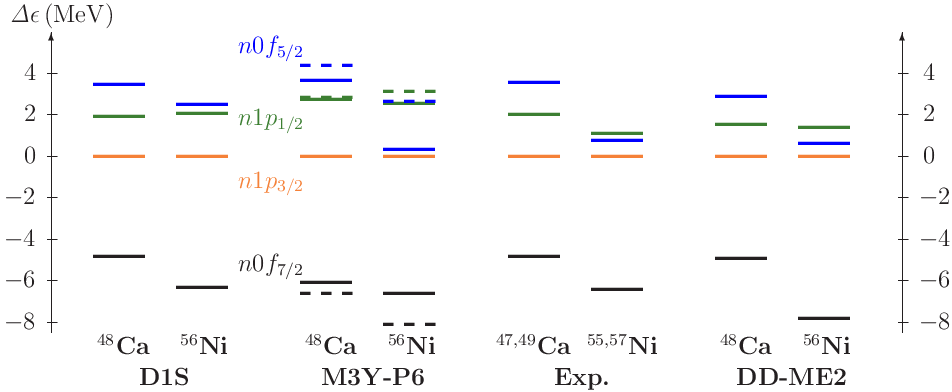}
  \caption{S.p. energy of $n0f_{5/2}$ (blue bars)
    measured from its PSS partner $n1p_{3/2}$ (orange bars)
    at $^{48}$Ca and $^{56}$Ni.
    The spherical HF results with D1S and M3Y-P6 are depicted.
    The result after subtracting the tensor-force contribution
    from that with M3Y-P6 is shown by the dashed bars,
    and the RMF result with DD-ME2 is also presented.
    Energies of $n0f_{7/2}$ (black bars) and $n1p_{1/2}$ (green bars)
    are also displayed for reference.
    For comparison, the measured energies of $^{49}$Ca and $^{57}$Ni
    of the lowest levels with corresponding spin-parities~\cite{ref:NuDat}
    are plotted.
    For $n0f_{7/2}$, the reference energy is extracted
    from the binding energies of $^{47}$Ca and $^{55}$Ni~\cite{ref:AME2020}.
    \label{fig:splevel_Ca48-Ni56}}
\end{figure}

We first look at the neutron shell structure above $N=28$.
The near degeneracy of $n0f_{5/2}$ and $n1p_{3/2}$ in the Ni nuclei
is a typical example of the PSS~\cite{ref:AHS69}.
On the other hand, the magicity of $N=32$ near $^{52}$Ca~\cite{ref:Ca52_Ex2}
occurs due to the large spacing between these orbits,
implying that the PSS significantly depends on the proton number $Z$.
In Fig.~\ref{fig:splevel_Ca48-Ni56},
the calculated s.p. energy spacing between the PSS partners at $^{48}$Ca
is compared with the spacing at $^{56}$Ni.
Occupation of protons on $p0f_{7/2}$
gives rise to the difference between $^{48}$Ca and $^{56}$Ni.
This $Z$-dependence of the shell structure
may largely be ascribed to the tensor force~\cite{ref:Nak10b}.
The tensor force is active at $^{56}$Ni owing to the occupation on $p0f_{7/2}$,
while its effect is small at $^{48}$Ca
because of the $\ell s$-closure for protons.
The $Z$-dependence in the M3Y-P6 result is compatible
with the experimental levels extracted from the lowest levels
at $^{49}$Ca and $^{57}$Ni.
In contrast, D1S does not give the $Z$-dependence
accounting for the PSS around the Ni nuclei.
If we subtract the tensor-force contribution,
M3Y-P6 reaches quite a similar result to that of D1S
as shown by the dashed bars in Fig.~\ref{fig:splevel_Ca48-Ni56},
confirming that the tensor force is responsible for the $Z$-dependence.
In all examples handled in this paper,
the s.p. spacings with M3Y-P6 resemble those with D1S
if subtracting the tensor-force contribution.

Intriguingly, the RMF result with DD-ME2 also describes the $Z$-dependence.
Without the exchange terms,
the origin of the $Z$-dependence in the DD-ME2 result
should be attributed to the central channel.
When we look at the calculated $n0f_{7/2}$ level
together with $n0f_{5/2}$ and $n1p_{3/2}$,
we find that both $n0f_{7/2}$ and $n0f_{5/2}$ shift down
from $^{48}$Ca to $^{56}$Ni in the DD-ME2 result,
almost keeping the spacing between them.
Thus, the $Z$-dependence of the PSS in the DD-ME2 result
originates from the different $Z$-dependence
among the s.p. levels with different $\ell$ values,
not greatly varying the $\ell s$ splitting.
This is a sharp contrast to the M3Y-P6 result.
Similar behavior is found in other regions shown below.
If we extract the s.p. energies from the measured binding energies
of $^{47}$Ca and $^{55}$Ni,
the variation of the $n0f_{7/2}$ level is
inbetween the M3Y-P6 and DD-ME2 results.
Considering together the ambiguity in extracting the s.p.energies
due to the fragmentation of the s.p. states,
these data cannot conclusively tell the preference
between the tensor-force picture and the relativistic picture
of the $Z$-dependence.

The relative position of $n1p_{1/2}$ is also displayed
in Fig.~\ref{fig:splevel_Ca48-Ni56}.
The $1/2^-$ level lies closely to $3/2^-$ and $5/2^-$ experimentally.
The position of the $1/2^-$ level does not look
to be well reproduced at $^{56}$Ni in the D1S and M3Y-P6 results.
However, our main concern is the $Z$-dependence of the level spacing.
The spacing between $n1p_{3/2}$ and $n1p_{1/2}$ does not change much
from $^{48}$Ca to $^{56}$Ni in all the results,
although it slightly widens in the D1S and tensor-subtracted M3Y-P6 results,
opposite to the experimental data.

\begin{figure}
  \includegraphics[scale=1.0]{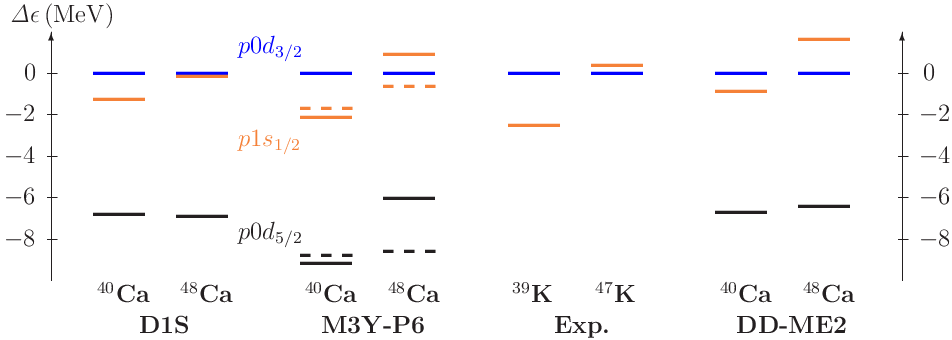}
  \caption{S.p. energy of $p1s_{1/2}$ (orange bars)
    measured from its PSS partner $p0d_{3/2}$ (blue bars)
    at $^{40}$Ca and $^{48}$Ca.
    Energy of $p0d_{5/2}$ (black bars) is also displayed.
    The measured energies of $^{39}$K and $^{47}$K~\cite{ref:NuDat}
    are shown for comparison.
    See Fig.~\protect\ref{fig:splevel_Ca48-Ni56} for other conventions.
    \label{fig:splevel_Ca40-Ca48}}
\end{figure}

Let us turn to other cases.
Whereas the $p1s_{1/2}$ level lies well below its PSS partner $p0d_{3/2}$
near $^{40}$Ca,
the two levels are inverted and closely lying around $^{48}$Ca,
giving an example of the $N$-dependence of the PSS.
It has been pointed out that the energy difference of these s.p. levels
is governed by the tensor force~\cite{ref:NSM13}.
The role of the tensor force is again confirmed
in Fig.~\ref{fig:splevel_Ca40-Ca48},
in which the variation of the energy spacing between $p0d_{3/2}$ and $p1s_{1/2}$
from $^{40}$Ca to $^{48}$Ca is shown.
The tensor-force effect is sizable at $^{48}$Ca
owing to the occupation on $n0f_{7/2}$,
while small at $^{40}$Ca.
The experimental levels are extracted from the lowest levels
of $^{39}$K and $^{47}$K with corresponding spin-parity.
The energy difference is similar to the one
obtained from the proton knockout reaction data,
which almost exhaust the spectroscopic factors~\cite{ref:DWKM76,ref:Ogi87}.
The variation of the difference between $p0d_{3/2}$ and $p1s_{1/2}$
is insufficient in the RMF result~\cite{ref:GMK07,ref:WGZD11}.
Notable difference is found between the full M3Y-P6 result and the others
in the variation of $p0d_{5/2}$ from $^{40}$Ca to $^{48}$Ca,
reflecting the tensor force,
though the energy of this deep hole state is not easy to evaluate
from experimental data because of its fragmentation.

\begin{figure}
  \includegraphics[scale=1.0]{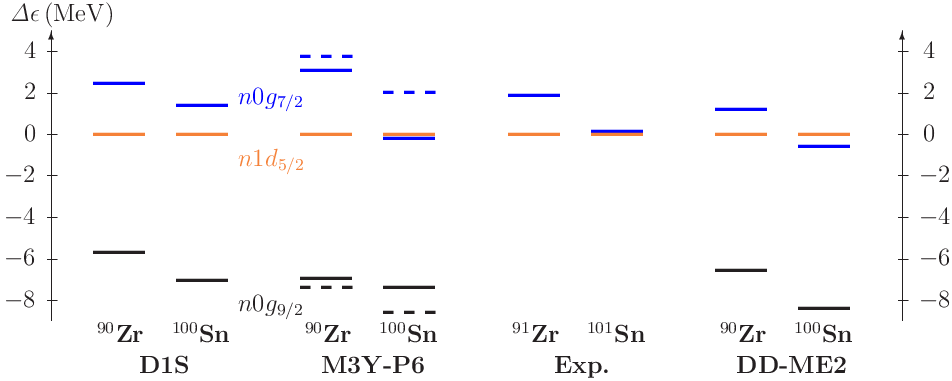}
  \caption{S.p. energy of $n0g_{7/2}$ (blue bars)
    measured from its PSS partner $n1d_{5/2}$ (orange bars)
    at $^{90}$Zr and $^{100}$Sn.
    Energy of $n0g_{9/2}$ (black bars) is also displayed.
    The measured energies of $^{91}$Zr and $^{101}$Sn~\cite{ref:NuDat}
    are shown for comparison.
    See Fig.~\protect\ref{fig:splevel_Ca48-Ni56} for other conventions.
    \label{fig:splevel_Zr90-Sn100}}
\end{figure}

The $E2$ excitation strengths in the neutron-deficient Sn nuclei
are hardly described
as far as the stiff spherical shape is assumed for their ground states~\cite{
  ref:Tog18,ref:ONAT23}.
The near degeneracy of the PSS partners $n0g_{7/2}$ and $n1d_{5/2}$
seems to trigger their deformability.
In contrast, $N=56$ may be a submagic number
at $^{96}$Zr~\cite{ref:NS14,ref:MN18},
whose excitation energy is relatively high~\cite{ref:NuDat}.
The $n0g_{7/2}$ level is distant from $n1d_{5/2}$ at Zr
but becomes close as $p0g_{9/2}$ is occupied,
providing another example of the $Z$-dependence of the PSS.
We investigate the $Z$-dependence of the s.p. level spacing
by picking up the $N=50$ nuclei.
As observed in Fig.~\ref{fig:splevel_Zr90-Sn100},
M3Y-P6 well describes the experimental $Z$-dependence,
owing to the tensor force.
The level spacing between $n0g_{7/2}$ and $n1d_{5/2}$
seems too large at Sn with D1S and too small at Zr with DD-ME2.
Although the variation of $n0g_{9/2}$ is distinguished
between M3Y-P6 and the others,
the data on $^{99}$Sn is required to access it experimentally,
which lies outside the proton-drip line.

\begin{figure}
  \includegraphics[scale=1.0]{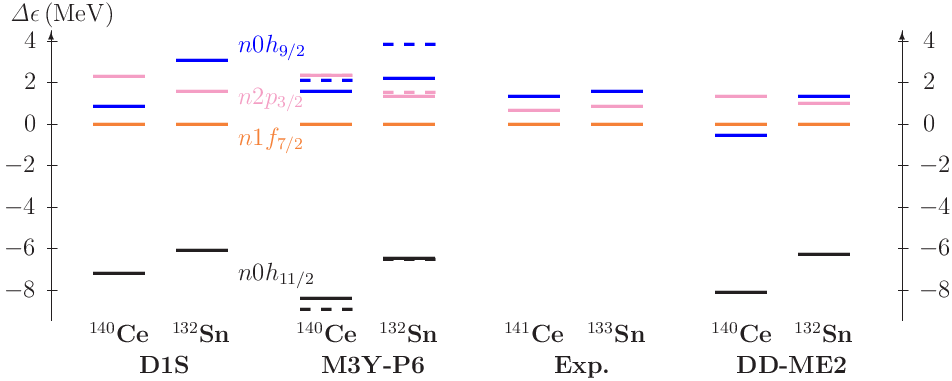}
  \caption{S.p. energy of $n0h_{9/2}$ (blue bars)
    measured from its PSS partner $n1f_{7/2}$ (orange bars)
    at $^{140}$Ce and $^{132}$Sn.
    Energies of $n0h_{11/2}$ (black bars) and $n2p_{3/2}$ (lavender bars)
    are also displayed.
    The measured energies of $^{141}$Ce and $^{133}$Sn~\cite{ref:NuDat}
    are shown for comparison.
    See Fig.~\protect\ref{fig:splevel_Ca48-Ni56} for other conventions.
    \label{fig:splevel_Ce140-Sn132}}
\end{figure}

Via the measurements of the isotope shifts,
the differential charge radii have been extracted with high precision.
Apart from exceptions like the neutron-deficient Sn nuclei mentioned above,
the proton configuration hardly changes in magic-$Z$ isotopes,
\textit{e.g.}, Pb, Sn, Ni and Ca.
It has been disclosed that the charge radii prevalently have kinks
at magic $N$~\cite{ref:Ang13,ref:shift-Ca52,ref:shift-Sn134,ref:Nak19}.
For Pb, the near degeneracy of the PSS partners $n0i_{11/2}$ and $n1g_{9/2}$
was argued to be one of the key ingredients providing the kink
at $^{208}$Pb~\cite{ref:SLKR95,ref:RF95,ref:GSR13,ref:NI15,ref:shift-Hg}.
The near degeneracy of $n0h_{9/2}$ and $n1f_{7/2}$
could be important~\cite{ref:Nak15}
in the newly discovered kink at $^{132}$Sn~\cite{ref:shift-Sn134}, as well.
It is noted that not all RMF parameter-sets successfully reproduce
the kink at $^{132}$Sn,
possibly related to the degree of the PSS~\cite{ref:NOSW23}.
In Fig.~\ref{fig:splevel_Ce140-Sn132},
the variation of the energy spacing between $n0h_{9/2}$ and $n1f_{7/2}$
from $^{140}$Ce to $^{132}$Sn is presented.
At $^{140}$Ce, $p0g_{7/2}$ is fully occupied in the calculated results,
making the tensor-force effect on the neutron orbitals vanishingly small.
As the observed excitation energy is relatively high~\cite{ref:NuDat},
$Z=58$ could be a submagic number at $^{140}$Ce~\cite{ref:NS14} in reality.
The $n0h_{9/2}$ orbit lies above $n1f_{7/2}$ at $^{140}$Ce,
and it further goes up at $^{132}$Sn.
However, the tensor force suppresses the increase of the $n0h_{9/2}$ energy,
keeping the PSS a good approximation and producing a kink at $^{132}$Sn
via a three-nucleon-force effect~\cite{ref:Nak15},
whereas another study ascribes the origin of the kink
to the property of the pairing~\cite{ref:Fay00}.
Analogously, the PSS maintained by the tensor force is important
for the kink of the charge radii at $^{208}$Pb.

\begin{figure}
  \includegraphics[scale=1.0]{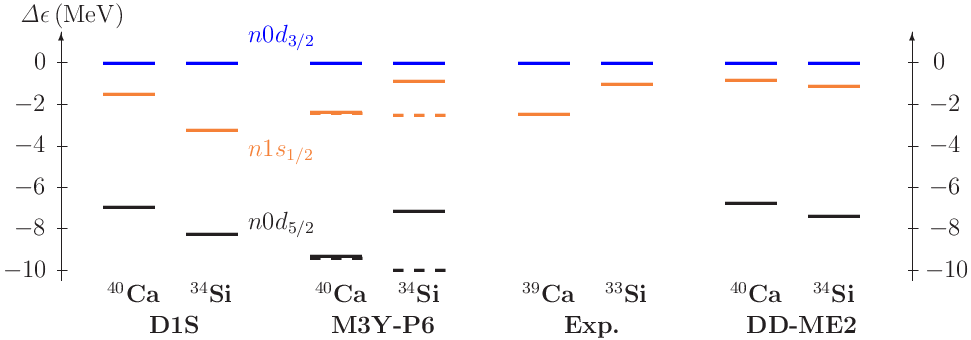}
  \caption{S.p. energy of $n1s_{1/2}$ (orange bars)
    measured from its PSS partner $n0d_{3/2}$ (blue bars)
    at $^{40}$Ca and $^{34}$Si.
    Energy of $n0d_{5/2}$ (black bars) is also displayed.
    The measured energies of $^{39}$Ca and $^{33}$Si~\cite{ref:NuDat}
    are shown for comparison.
    See Fig.~\protect\ref{fig:splevel_Ca48-Ni56} for other conventions.
    \label{fig:splevel_Ca40-Si34}}
\end{figure}

Let us go back to the region near $^{40}$Ca.
The $n1s_{1/2}$ orbit lies below its PSS partner $n0d_{3/2}$.
It may go up and down as $p0d_{3/2}$ becomes unoccupied.
D1S and DD-ME2, which do not contain explicit tensor force,
predict $n1s_{1/2}$ goes down from $^{40}$Ca to $^{34}$Si,
while M3Y-P6 gives the opposite trend.
These qualitatively different predictions are noteworthy,
because they seem distinctive of the tensor-force effect.
The $^{34}$Si nucleus is expected to have a doubly-magic nature,
with a good possibility of the proton semi-bubble density~\cite{ref:NSM13,
  ref:LLSZ16}.
Therefore, experimental data at $^{39}$Ca and $^{33}$Si could supply a reference.
As shown in Fig.~\ref{fig:splevel_Ca40-Si34},
the measured energies of the lowest levels are consistent
with the M3Y-P6 result.

We have discussed variations of the s.p. level spacings
between the pseudo-spin partners
through the occupation of specific orbits.
One might expect that a long chain of isotopes or isotones
gives clear indication of tensor-force effects in the PSS,
possibly discriminating them from the relativistic effects.
In Figs.~14, 15 and 16 of Ref.~\cite{ref:Nak08b},
the variation of the s.p. level spacing was investigated
for the chains of $Z=50$, $N=50$ and $N=82$ nuclei,
respectively.
To elucidate the roles of the tensor force,
results of the non-relativistic interactions
with and without tensor force were compared.
However, the central channels in the interaction,
which are more or less adjusted to experimental data,
also influence the level spacing,
and obscure the tensor-force effects,
whereas the tensor force significantly influences
the relative energies of the unique-parity orbits.
We have confirmed that this holds for the DD-ME2 case.

\textit{Summary and discussion.}\quad

We have pointed out the relevance of
the $Z$- and $N$-dependence of the shell structure,
often called ``shell evolution'',
to the pseudo-spin symmetry (PSS).
This indicates that the tensor force may play a significant role in the PSS.
We have investigated the tensor-force effects on the PSS
by employing the spherical Hartree-Fock calculations with M3Y-P6.
It should be noted that the tensor force is undoubtedly contained
in the nucleonic interaction,
and the tensor force adopted here is a realistic one
derived from the $G$-matrix.
We find that the tensor-force effects on the PSS
sometimes look analogous to the $Z$-and $N$-dependence of the PSS
in the relativistic mean-field (RMF) calculations without explicit tensor force,
despite difference in the physics origin.
However, for the variation of the $p0d_{3/2}$-$p1s_{1/2}$ levels
from $^{40}$Ca to $^{34}$Si,
the experimental data is compatible with the tensor-force-driven picture
of the PSS,
but not with the prediction of the RMF with the DD-ME2 Lagrangian.
Although it is too early to conclude from this case
that the tensor force dominates the $Z$- and $N$-dependence of the PSS,
it demonstrates that the tensor force may be significant in the PSS.
Conversely, it could be misleading to discuss the PSS
only within the context of relativistic symmetry
without paying attention to the tensor-force effects.

The tensor force between nucleons arises
in the relativistic Hartree-Fock (RHF) framework~\cite{ref:BMGM87,ref:Long07},
which takes account of the exchange terms.
The role of the tensor coupling due to the $\rho$-meson in the PSS
has been argued~\cite{ref:Geng19},
which gives a part of the tensor-force effect discussed here.
See Ref.~\cite{ref:WZLL18}
for the relation of the meson-nucleon coupling in the RHF
to the nucleonic tensor force.
As the tensor force and the relativistic effects may affect the PSS
cooperatively and competitively,
it is of interest to view how the $Z$- and $N$-dependence of the PSS
arises within the RHF scheme.

~

Discussions with N.~Hinohara and T.~Naito are gratefully acknowledged.
This work is supported by the JSPS KAKENHI, Grant No. JP24K07012.
A part of the numerical calculations has been performed on HITAC SR24000
at Institute of Management and Information Technologies at Chiba University.


\end{document}